\documentclass{emulateapj}

\slugcomment{Accepted for publication in \textit{The Astrophysical Journal}}
\usepackage[colorlinks,urlcolor=blue,citecolor=blue,linkcolor=blue]{hyperref} 
\usepackage{graphicx}
\usepackage{color}

\shorttitle{Low Luminosity Galaxy Clusters}
\shortauthors{Connor et al.}

\begin{document}

\newcommand{\Lx}{$\mathrm{L}_\mathrm{X}$}
\newcommand{\Tx}{$\mathrm{T}_\mathrm{X}$}
\newcommand{\Mx}{$\mathrm{M}_\mathrm{X}$}
\newcommand{\Mrad}{$\mathrm{M}_{2500}$}
\newcommand{\rrad}{$\mathrm{r}_{2500}$}
\newcommand{\Msol}{$\mathrm{M}_\odot$}

\title{Scaling Relations and X-ray Properties of Moderate-Luminosity Galaxy Clusters from $0.3 < \MakeLowercase{z} < 0.6$ with \textit{XMM-Newton}\footnotemark[1]}
\footnotetext[1]{Based on observations obtained with \textit{XMM-Newton}, an ESA science mission with instruments and contributions directly funded by ESA Member States and NASA}

\author{Thomas Connor\altaffilmark{2,3}, Megan Donahue\altaffilmark{2}, Ming Sun\altaffilmark{4}, Henk Hoekstra\altaffilmark{5}, Andisheh Mahdavi\altaffilmark{6}, Christopher J. Conselice\altaffilmark{7}, and Brian McNamara\altaffilmark{8,9,10} }

\altaffiltext{2}{Department of Physics and Astronomy, Michigan State University, East Lansing, MI 48823, USA}
\altaffiltext{3}{connort1@msu.edu}
\altaffiltext{4}{Department of Physics, University of Alabama in Huntsville, Huntsville, AL 35899, USA}
\altaffiltext{5}{Leiden Observatory, Leiden University, PO Box 9513, 2300 RA, Leiden, the
Netherlands}
\altaffiltext{6}{Department of Physics and Astronomy, San Francisco State University, San Francisco, CA 94131, USA}
\altaffiltext{7}{University of Nottingham, School of Physics and Astronomy, Nottingham NG7 2RD, UK}
\altaffiltext{8}{Department of Physics and Astronomy, University of Waterloo, 200 University Avenue West, Waterloo, Ontario N2L 5M3, Canada}
\altaffiltext{9}{Perimeter Institute for Theoretical Physics, 31 Caroline Street, Waterloo, Ontario N2L 2Y5, Canada}
\altaffiltext{10}{Harvard-Smithsonian Center for Astrophysics, Cambridge, MA 02138, USA}
\begin{abstract}
We present new X-ray temperatures and improved X-ray luminosity estimates for 15 new and archival \textit{XMM-Newton} observations of galaxy clusters at intermediate redshift with mass and luminosities near the galaxy group/cluster division ($ \mathrm{M}_{2500}  < 2.4 \times 10^{14} \, h_{70}^{-1}\,\mathrm{M}_\odot$, $\mathrm{L} < 2 \times 10^{44}\, \mathrm{erg} \, \mathrm{s}^{-1}$, $0.3 < \mathrm{z} < 0.6$). These clusters have weak-lensing mass measurements based on \textit{Hubble Space Telescope} observations of clusters representative of an X-ray selected sample (the \textit{ROSAT} 160SD survey). The angular resolution of \textit{XMM-Newton} allows us to disentangle the emission of these galaxy clusters from nearby point sources, which significantly contaminated previous X-ray luminosity estimates for six of the fifteen clusters. We extend cluster scaling relations between X-ray luminosity, temperature, and weak-lensing mass for low-mass, X-ray-selected clusters out to redshift $\sim0.45$. These relations are important for cosmology and the astrophysics of feedback in galaxy groups and clusters. Our joint analysis with a sample of 50 clusters in a similar redshift range but with larger masses ($ \mathrm{M}_{500}  < 21.9 \times 10^{14} \,\mathrm{M}_\odot$, $0.15 \leq \mathrm{z} \leq 0.55$) from the Canadian Cluster Comparison Project finds that within $\mathrm{r}_{2500}$, $\mathrm{M} \propto \mathrm{L}^{0.44 \pm 0.05}$, $\mathrm{T} \propto \mathrm{L}^{0.23 \pm 0.02}$, and $\mathrm{M} \propto \mathrm{T}^{1.9 \pm 0.2}$. The estimated intrinsic scatter in the M-L relation for the combined sample is reduced to $\sigma_{\log (M|L)} = 0.10$, from $\sigma_{\log (M|L)} = 0.26$ with the original \textit{ROSAT} measurements. We also find an intrinsic scatter for the T-L relation, $\sigma_{\log (T|L)} = 0.07 \pm 0.01.$
\end{abstract}


\section{Introduction}

Simulations of cosmological structure formation show clusters and filaments of dark matter growing from a set of random initial perturbations into a cosmic web \citep[e.g.][]{2005Natur.435..629S, 2009MNRAS.398.1150B, 2011ApJ...740..102K}. The statistical properties of this cosmic web are extremely sensitive to the values of certain cosmological parameters, particularly $\Omega_M$ and $\sigma_8$, the amplitude of the initial perturbation spectrum \citep[e.g][]{1996MNRAS.282..263E, 1998ApJ...504....1B, 2001ApJ...560L.111H, 2011ARAA..49..409A}.

This cosmic web of dark matter is easiest to investigate by studying its most massive systems, which are clusters of galaxies. About 85\% of a cluster's mass is composed of dark matter, while nearly all of the rest is intergalactic hot gas, with a trace amount contributed by stars \citep{2002ARAA..40..539R, 2005RvMP...77..207V, 2006ApJ...652..917L}.  The hot gas is confined by the cluster's gravitational potential and radiates X-rays, providing powerful diagnostics for properties of the host cluster, including its mass, baryonic content, and dynamic status.

Accurate measurements of galaxy cluster masses are useful for more than just describing individual systems; galaxy cluster masses are needed to verify models of large-scale structure formation \citep{2001MNRAS.321..372J, 2007MNRAS.382.1261G, 2009ApJ...692.1033V} and to constrain cosmological parameters \citep{2008ApJ...688..709T, 2010ApJ...708..645R, 2010MNRAS.406.1759M, 2011ApJ...732..122B}. One way to accurately measure the projected mass of a cluster is through measurements of gravitational lensing \citep{2013SSRv..177...75H}. However, performing such measurements is prohibitively expensive for large samples of clusters and difficult for low redshift clusters. To this end, scaling relations have been empirically calibrated to connect observed properties to masses. Examples of this include $\mathrm{L}_\mathrm{X} - \mathrm{M}_\mathrm{X}$, $\mathrm{M}_\mathrm{X} - \mathrm{T}_\mathrm{X}$, and $\mathrm{M}_\mathrm{X} - \mathrm{Y}_\mathrm{X}$ relations.

Early work by \citet{1986MNRAS.222..323K} showed that these relations can be cast analytically for the case where cold gas falls into preexisting dark matter structures. Those early relations predicted clusters that were overluminous for a given temperature compared to observations. So \citet{1991ApJ...383..104K} and \citet{1991ApJ...383...95E} showed that preheating could increase the entropy of intergalactic gas. Such a ``preheating'' model elevates the entropy of the gas, preventing the gas from getting too dense. These predicted 
$\mathrm{L}_\mathrm{X} - \mathrm{M}_\mathrm{X}$ and $\mathrm{T}_\mathrm{X} - \mathrm{M}_\mathrm{X}$ relations were roughly consistent with observations. This expectation that clusters would follow such laws over a large range of M$_X$, with standard evolutionary factors, is known as self-similarity \citep{1997ApJ...490..493N, 1997MNRAS.288..355B, 1998ApJ...495...80B}. The scale-free nature of this behavior arises because the gravitational potential dominates over other energy sources, and gravity is scale-free.

Previous work has shown possible deviations from self-similarity at masses approaching those of galaxy groups \citep[e.g.][]{1996MNRAS.283..690P, 2000ApJ...538...65X, 2011AA...535A.105E, 2012MNRAS.422.2213S}, possibly due to the increasing fractional contribution of local feedback processes to the cluster energy budget compared to the gravitational potential. The exact magnitude and behavior of this deviation is not yet defined, but it has been qualitatively reproduced in numerical work \citep{2008ApJ...687L..53P, 2010MNRAS.401.1670F}. 

A full understanding of the deviation from self-similarity can only come through a thorough exploration of the cluster parameter space -- across cluster mass ranges and redshifts. One under-sampled regime is at moderate redshift and low mass. Clusters with these properties offer us the ability to answer the questions of how scaling relations change from high redshift to low redshift and whether there is any evolution in the low-mass behavior of these relations. 

Recent work by \citet{2011ApJ...726...48H} provided weak lensing mass measurements from the \textit{Hubble Space Telescope} (HST) of 25 galaxy clusters occupying this redshift regime. That work lacked high-quality X-ray observations for most of the objects, however. We use observations with the \textit{XMM-Newton} satellite to study the X-ray characteristics of this sample and to constrain X-ray property and mass scaling relations for this redshift and mass regime.

The structure of this paper is as follows. In Section \ref{sect:Data&Analysis}, we describe the properties of our sample, while our analysis techniques are described in Section \ref{sect:Analysis}. The results of our analysis are presented in Section \ref{sect:Results}. In particular, we discuss our fits of three scaling relations involving X-ray luminosity, temperature, and weak lensing mass. Finally, we compare our results to other published works in Section \ref{sect:Discussion} and Appendix \ref{apx:xmm_comp}. Throughout this paper, we assume a flat $\Lambda$CDM cosmology with $\Omega_M = 0.3$, and $H_0 = 70\, \mathrm{km}\, \mathrm{s}^{-1}\,\mathrm{Mpc}^{-1}$.

\section{Data and Analysis}
\label{sect:Data&Analysis}
\begin{deluxetable*}{lllccccc}
\tabletypesize{\scriptsize}
\tablecaption{Sample Properties}
\tablewidth{0pt}
\tablehead{
\colhead{Cluster Name} & \colhead{$\alpha_{2000}$\tablenotemark{a}} & \colhead{$\delta_{2000}$\tablenotemark{a}} & \colhead{$\alpha_{2000}$\tablenotemark{b}} & \colhead{$\delta_{2000}$\tablenotemark{b}} &\colhead{z\tablenotemark{c}} & \colhead{$N_H$\tablenotemark{d}}  & \colhead{$r_{2500}$\tablenotemark{a}}  \\
\colhead{} & \colhead{} & \colhead{} & \colhead{} & \colhead{} & \colhead{} &  \colhead{$10^{20}$ $\mathrm{cm}^{-2}$} & \colhead{$h_{70}^{-1}$ Mpc}
} 
\startdata
RXJ0056.9$-$2740 & $00^\mathrm{h}56^\mathrm{m}56.98^\mathrm{s}$ & $-27^\circ40^\mathrm{m}29.9^\mathrm{s}$ & $00^\mathrm{h} 56^\mathrm{m} 57.9^\mathrm{s}$ & $ -27^\circ 40^\mathrm{m} 29.3^\mathrm{s} $ & 0.563 &  1.79 & 0.270 \\ 
RXJ0110.3+1938 & $01^\mathrm{h}10^\mathrm{m}18.22^\mathrm{s}$ & $+19^\circ38^\mathrm{m}19.4^\mathrm{s}$ & $01^\mathrm{h} 10^\mathrm{m} 18.2^\mathrm{s}$ & $ +19^\circ 38^\mathrm{m} 18.7^\mathrm{s} $ & 0.317 & 3.82  & 0.293 \\ 
RXJ0522.2$-$3625 & $05^\mathrm{h}22^\mathrm{m}15.48^\mathrm{s}$ & $-36^\circ24^\mathrm{m}56.1^\mathrm{s}$ & $05^\mathrm{h} 22^\mathrm{m} 15.4^\mathrm{s}$ & $ -36^\circ 24^\mathrm{m} 55.7^\mathrm{s} $ & 0.472 &  3.63 & 0.313 \\ 
RXJ0826.1+2625 & \nodata & \nodata & $08^\mathrm{h}26^\mathrm{m}08.03^\mathrm{s}$ & $+26^\circ25^\mathrm{m}16.7^\mathrm{s}$ & 0.351 & 3.39  & 0.157 \\ 
RXJ0847.1+3449 & $08^\mathrm{h}47^\mathrm{m}11.79^\mathrm{s}$ & $+34^\circ48^\mathrm{m}51.8^\mathrm{s}$ & $08^\mathrm{h} 47^\mathrm{m} 11.7^\mathrm{s}$ & $ +34^\circ 48^\mathrm{m} 51.9^\mathrm{s} $ & 0.560 & 2.92  & 0.452 \\ 
RXJ0957.8+6534 & $09^\mathrm{h}57^\mathrm{m}51.22^\mathrm{s}$ & $+65^\circ34^\mathrm{m}25.1^\mathrm{s}$ & $09^\mathrm{h} 57^\mathrm{m} 51.1^\mathrm{s}$ & $ +65^\circ 34^\mathrm{m} 26.1^\mathrm{s} $ & 0.530 & 5.32  & 0.257 \\ 
RXJ1117.4+0743 & $11^\mathrm{h}17^\mathrm{m}26.04^\mathrm{s}$ & $+07^\circ43^\mathrm{m}38.3^\mathrm{s}$ & $11^\mathrm{h} 17^\mathrm{m} 26.1^\mathrm{s}$ & $ +07^\circ 43^\mathrm{m} 41.0^\mathrm{s} $ & 0.477 & 3.59  & 0.280 \\ 
RXJ1354.2$-$0221 & $13^\mathrm{h}54^\mathrm{m}17.19^\mathrm{s}$ & $-02^\circ21^\mathrm{m}59.0^\mathrm{s}$ & $13^\mathrm{h} 54^\mathrm{m} 17.2^\mathrm{s}$ & $ -02^\circ 21^\mathrm{m} 59.4^\mathrm{s} $ & 0.546 & 3.22  & 0.428 \\ 
RXJ1642.6+3935 & $16^\mathrm{h}42^\mathrm{m}38.35^\mathrm{s}$ & $+39^\circ36^\mathrm{m}10.4^\mathrm{s}$ & $16^\mathrm{h} 42^\mathrm{m} 38.4^\mathrm{s}$ & $ +39^\circ 36^\mathrm{m} 07.9^\mathrm{s} $ & 0.355 & 1.20  & 0.239 \\ 
RXJ2059.9$-$4245 & $20^\mathrm{h}59^\mathrm{m}54.92^\mathrm{s}$ & $-42^\circ45^\mathrm{m}32.1^\mathrm{s}$ & $20^\mathrm{h} 59^\mathrm{m} 54.9^\mathrm{s}$ & $ -42^\circ 45^\mathrm{m} 34.8^\mathrm{s} $ & 0.323 & 3.13  & 0.280 \\ 
RXJ2108.8$-$0516 & $21^\mathrm{h}08^\mathrm{m}51.17^\mathrm{s}$ & $-05^\circ16^\mathrm{m}58.4^\mathrm{s}$ & $21^\mathrm{h} 08^\mathrm{m} 51.2^\mathrm{s}$ & $ -05^\circ 16^\mathrm{m} 57.6^\mathrm{s} $ & 0.319 & 6.30  & 0.210 \\ 
RXJ2139.9$-$4305 & $21^\mathrm{h}39^\mathrm{m}58.22^\mathrm{s}$ & $-43^\circ05^\mathrm{m}13.9^\mathrm{s}$ & $21^\mathrm{h} 39^\mathrm{m} 58.3^\mathrm{s}$ & $ -43^\circ 05^\mathrm{m} 14.2^\mathrm{s} $ & 0.376 & 1.63  & 0.292 \\ 
RXJ2146.0+0423 & $21^\mathrm{h}46^\mathrm{m}05.52^\mathrm{s}$ & $+04^\circ23^\mathrm{m}14.3^\mathrm{s}$ & $21^\mathrm{h} 46^\mathrm{m} 05.6^\mathrm{s}$ & $ +04^\circ 23^\mathrm{m} 02.6^\mathrm{s} $ & 0.531 & 4.82  & 0.436 \\ 
RXJ2202.7$-$1902 & $22^\mathrm{h}02^\mathrm{m}45.50^\mathrm{s}$ & $-19^\circ02^\mathrm{m}21.1^\mathrm{s}$ & $22^\mathrm{h} 02^\mathrm{m} 45.5^\mathrm{s}$ & $ -19^\circ 02^\mathrm{m} 20.1^\mathrm{s} $ & 0.438 & 2.44  & 0.152 \\ 
RXJ2328.8+1453 & $23^\mathrm{h}28^\mathrm{m}52.27^\mathrm{s}$ & $+14^\circ52^\mathrm{m}42.8^\mathrm{s}$ & $23^\mathrm{h} 28^\mathrm{m} 52.3^\mathrm{s}$ & $ +14^\circ 52^\mathrm{m} 42.7^\mathrm{s} $ & 0.497 & 3.88  & 0.254
\enddata 
\tablenotetext{a}{Coordinates and $r_{2500}$ from \cite{2011ApJ...726...48H}.}\
\tablenotetext{b}{Coordinates from XMM centroid (see Section \ref{sect:xrayoffsets}).}\
\tablenotetext{c}{Cluster redshift from \cite{2003ApJ...594..154M}.}
\tablenotetext{d}{Column density from \cite{2005AA...440..775K}.} 
\label{tab:cluster_properties}
\end{deluxetable*}

\begin{deluxetable*}{llrrrr}
\tabletypesize{\scriptsize}
\tablecaption{Observations of Clusters}
\tablewidth{0pt}
\tablehead{
\colhead{Cluster Name} & \colhead{OBSID} & \colhead{Exposure Time} & \multicolumn{3}{c}{Usable Exposure Time}\\ 
\colhead{ } &\colhead{ } & \colhead{(s)} & \colhead{MOS1} & \colhead{MOS2} & \colhead{pn} }
\startdata
RXJ0056.9$-$2740 & 0111282001 &8876 & 8190 & 8017 & 4135 \\
RXJ0110.3+1938 & 0500940101 &32818 &18883 & 18497 & 6973 \\
RXJ0522.2$-$3625 & 0065760201 &31919 &31333 & 31317 & 26904 \\
\nodata & 0302580901 &31110 &20077 & 20284 & 16481 \\
RXJ0826.1+2625 & 0691670201\tablenotemark{a} &48742 &31516 & 31419 & 23135 \\
\nodata & 0603500301\tablenotemark{a} &40509 &19967 & 20463 &\nodata\\
RXJ0847.1+3449 & 0107860501 &91419 &58708 & 58333 &\nodata\\
RXJ0957.8+6534 & 0502430201 &72070 &44762 & 45062 & 30090 \\
RXJ1117.4+0743 & 0203560401 &86515 &81073 &\nodata& 56293 \\
\nodata & 0203560201 &81913 &71366 &\nodata& 57255 \\
\nodata & 0082340101 &63206 &60889 &\nodata& 43232 \\
RXJ1354.2$-$0221 & 0112250101 &33646 &24584 & 24000 &\nodata\\
RXJ1642.6+3935 & 0603500701\tablenotemark{a} &23917 &17108 & 17133 & 11099 \\
RXJ2059.9$-$4245 & 0691670101\tablenotemark{a} &57915 &56794 & 56571 & 41238 \\
RXJ2108.8$-$0516 & 0110860101 &38116 &34637 & 34668 &\nodata\\
RXJ2139.9$-$4305 & 0603501001\tablenotemark{a} &41916 &36715 & 36875 & 19382 \\
RXJ2146.0+0423 & 0302580701 &47120 &24091 & 24081 & 18316 \\
RXJ2202.7$-$1902 & 0203450201 &64117 &27842 & 26081 & 6919 \\
RXJ2328.8+1453 & 0502430301 &104910&94004 & 94249 & 70516
\enddata
\tablenotetext{a}{New data.}
\label{tab:OBSIDS}
\end{deluxetable*}

\begin{deluxetable}{ccc}
\tabletypesize{\scriptsize}
\tablecaption{Masked Sources}
\tablewidth{0pt}
\tablehead{
\colhead{$\alpha_{2000}$} & \colhead{$\delta_{2000}$} & \colhead{Radius} 
\\
\colhead{ } & \colhead{ } & \colhead{$('')$} 
}
\startdata
$00^\mathrm{h}56^\mathrm{m}48.44^\mathrm{s}$ & $-27^\circ40^\mathrm{m}59.5^\mathrm{s}$ & 12.9 \\
$00^\mathrm{h}56^\mathrm{m}49.65^\mathrm{s}$ & $-27^\circ40^\mathrm{m}07.6^\mathrm{s}$ & 25.8 \\
$00^\mathrm{h}57^\mathrm{m}04.10^\mathrm{s}$ & $-27^\circ41^\mathrm{m}11.5^\mathrm{s}$ & 21.5 \\
$00^\mathrm{h}57^\mathrm{m}04.70^\mathrm{s}$ & $-27^\circ40^\mathrm{m}23.5^\mathrm{s}$ & 21.5 \\
$00^\mathrm{h}57^\mathrm{m}09.22^\mathrm{s}$ & $-27^\circ39^\mathrm{m}39.5^\mathrm{s}$ & 20.5 
\enddata 
\tablecomments{Table \ref{tab:masked_source} is published in its entirety in the electronic edition of the Astrophysical Journal. A portion is shown here for guidance regarding its form and content.}
\label{tab:masked_source}
\end{deluxetable}

Our sample is based on 25 galaxy clusters first detected in the \textit{ROSAT} 160 Square Degree Survey. \cite{1998ApJ...502..558V} describe the initial survey, and a reanalysis with spectroscopic redshifts comes from \cite{2003ApJ...594..154M}. These 25 clusters were further studied with an HST snapshot program (PI: Donahue) of one orbit per cluster with the F814W filter. Due to the nature of the snapshot program, the clusters were randomly selected from a master list of 72 clusters. \cite{2011ApJ...726...48H} used those images to estimate weak-lensing masses for these clusters. The focus of this work is to improve and augment the X-ray measurements of these clusters with observations with \textit{XMM-Newton}. Along with new observations, we used archival data to supplement the cluster sample with new uniform measurements of X-ray properties.

We searched the archive of \textit{XMM-Newton} observations with the XMM-Newton Science Archive (XSA) v7.2 within a 15' radius of the cluster positions given in \cite{2011ApJ...726...48H}. As of June 28, 2013, we found 27 observations that included the cluster. We excluded 9 because they were too short and excluded 4 that were unusable due to excessive particle contamination from flares, leaving 14 observations of 11 clusters. We supplemented these with five new observations of four clusters. All observations were taken with the European Photon Imaging Camera (EPIC), which consists of two MOS cameras \citep{2001AA...365L..27T} and the pn camera \citep{2001AA...365L..18S}. Cluster properties drawn from earlier works are provided in Table \ref{tab:cluster_properties}. Hydrogen column density, $\mathrm{N}_\mathrm{H}$, is taken from the compilation by \cite{2005AA...440..775K}. The datasets used in this work are listed in Table \ref{tab:OBSIDS}. 

\begin{figure*}
	\begin{minipage}{0.5\textwidth}
		\includegraphics[width=\textwidth]{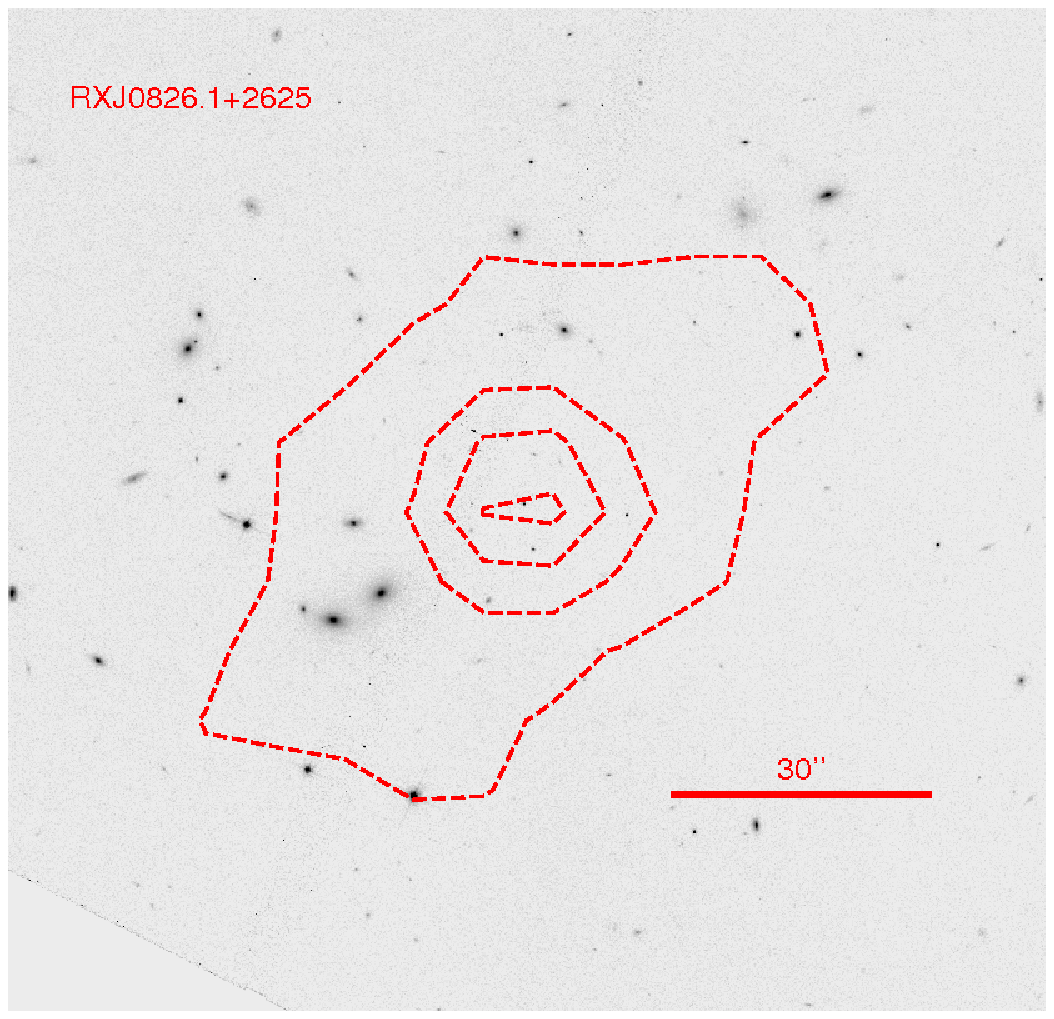}
    \end{minipage}
	\begin{minipage}{0.5\textwidth}
		\includegraphics[width=\textwidth]{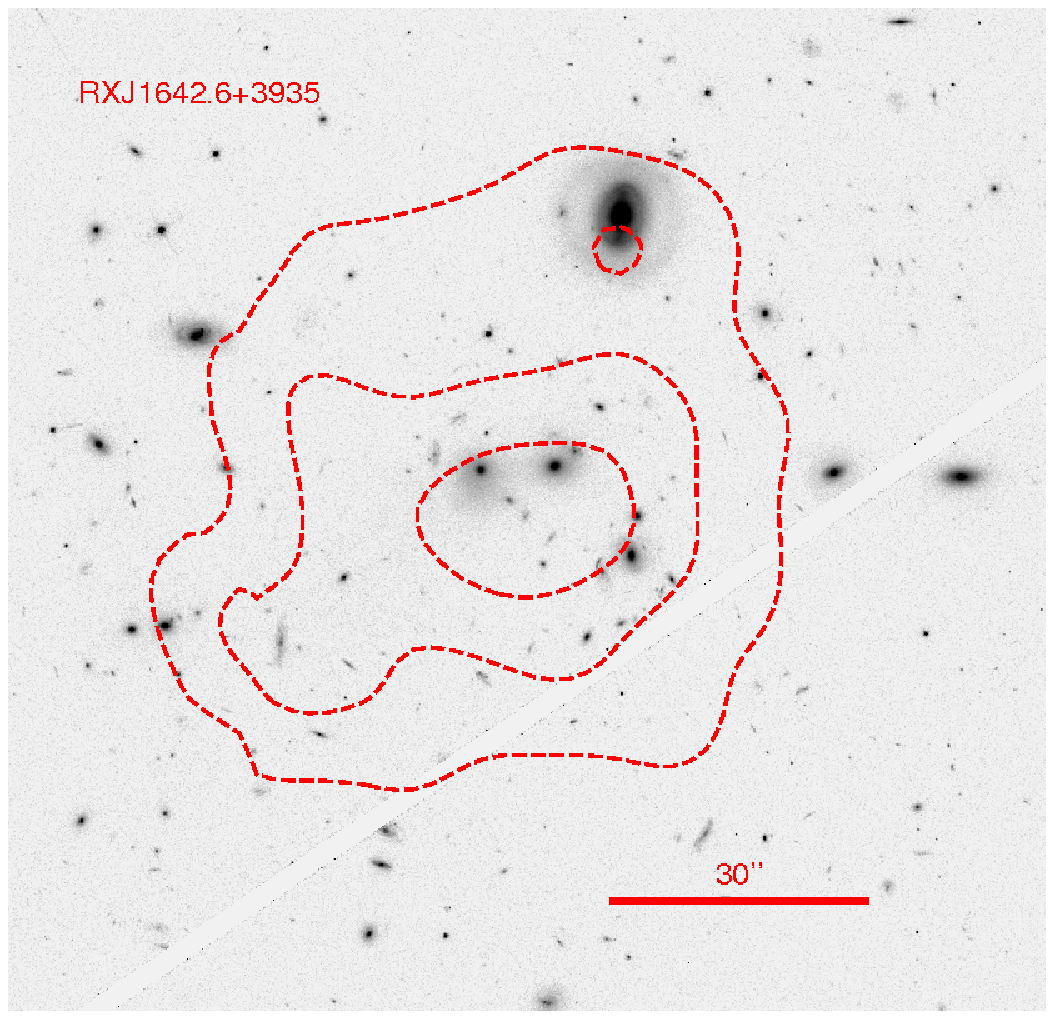}
    \end{minipage}
	\begin{minipage}{0.5\textwidth}
		\includegraphics[width=\textwidth]{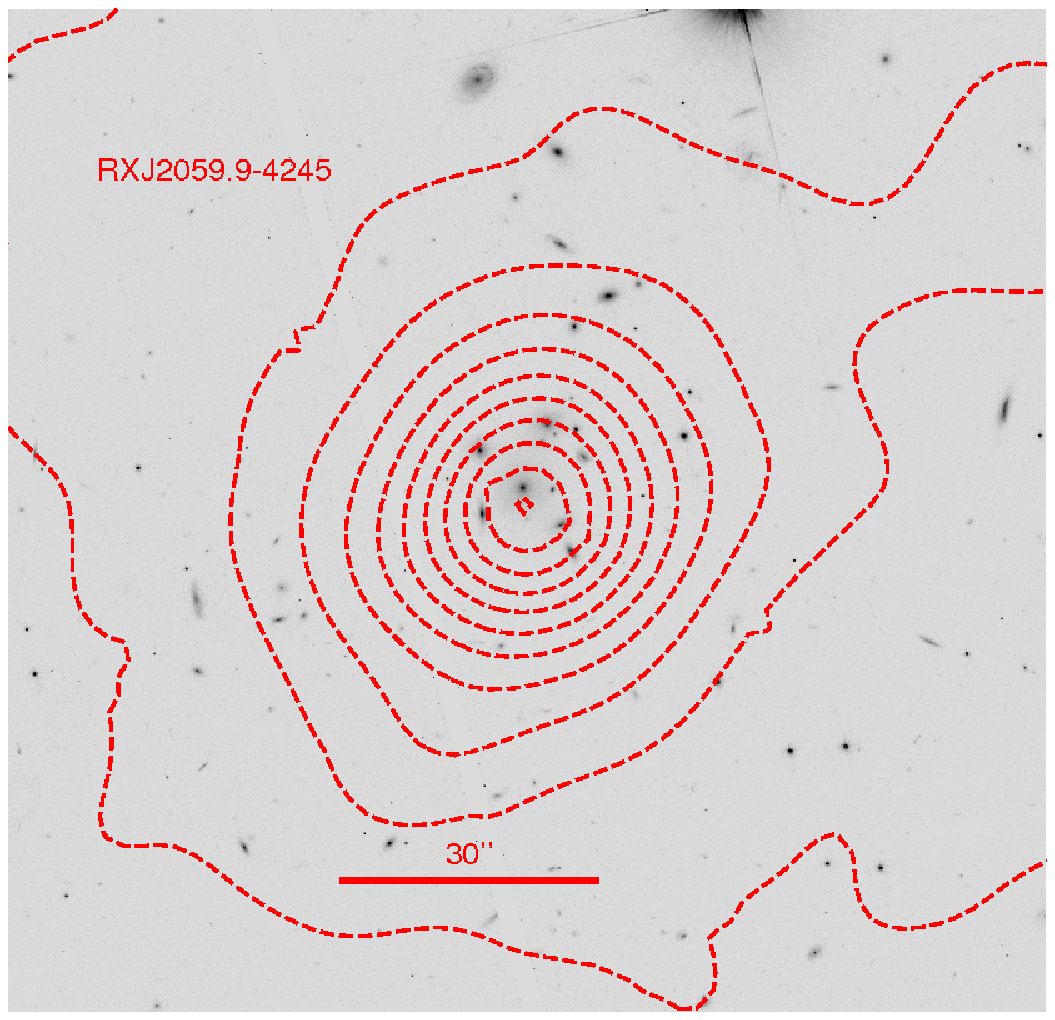}
    \end{minipage}
	\begin{minipage}{0.5\textwidth}
		\includegraphics[width=\textwidth]{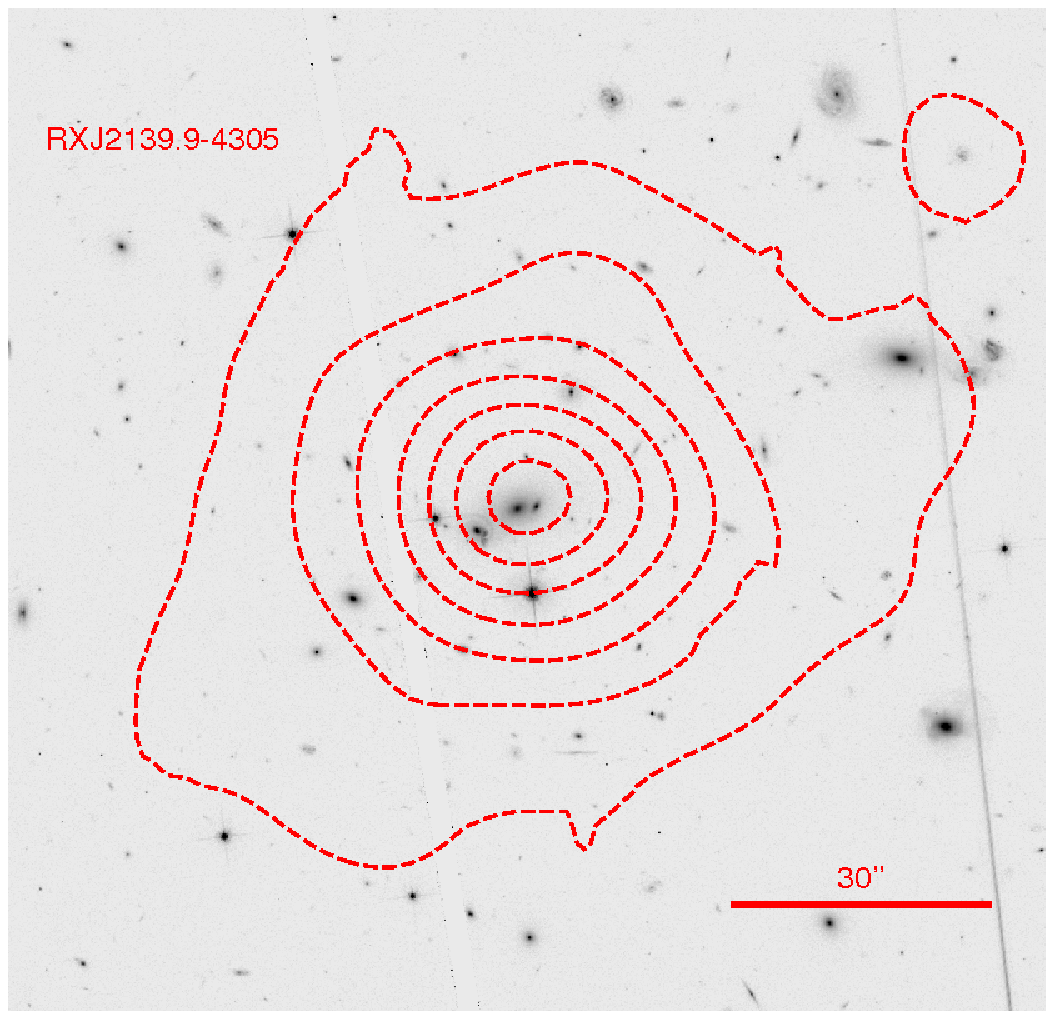}
    \end{minipage}
    \caption{Gaussian-smoothed X-ray emission contours overlaid on \textit{Hubble Space Telescope} images of the four clusters we observed in this work. Contours are spaced at intervals of $10^{-6} \, \mathrm{count}\, \mathrm{ s}^{-1}\,\mathrm{arcsec}^{-2}$, with the minimum level for each cluster described in Section \ref{sect:Data&Analysis}. }
\label{fig:new_obs}
\end{figure*}

Our new observations are presented in Figure \ref{fig:new_obs}. Our data are shown as smoothed X-ray contours from combined EPIC images overlaid on HST images of the cluster using the Advanced Camera for Surveys / Wide Field Channel F814W filter. Combined X-ray products were created using the XMM-Newton Science Analysis System (SAS) \texttt{images} script binning to 2$''$ and smoothing with a Gaussian FWHM of 15$''$ in the energy range of 0.4 - 8.0 keV. Contours are levels of $10^{-6} \, \mathrm{count}\, \mathrm{ s}^{-1}\,\mathrm{arcsec}^{-2}$, with the lowest displayed contour corresponding to $10^{-6} \, \mathrm{count}\, \mathrm{ s}^{-1}\,\mathrm{arcsec}^{-2}$ for RXJ0826.1+2625 and RXJ2059.9$-$4245 and $2 \times 10^{-6} \, \mathrm{count}\, \mathrm{ s}^{-1}\,\mathrm{arcsec}^{-2}$ for RXJ1642.6+3935 and RXJ2139.9$-$4305.

Observations were reduced using the XMM-Newton SAS version 12.0.1.\footnote{http://xmm.esac.esa.int/sas/} Bad time intervals were defined based on the count rate of high energy events (\textgreater 10 keV) in 100-second bins; time periods where those exceeded 0.35 count $\mathrm{s}^{-1}$ (MOS) or 0.40 count $\mathrm{s}^{-1}$ (pn) were excluded. One exception to this was the observation of RXJ1354.2$-$0221, 0112250101, which had an abnormally high high-energy background. To avoid overfiltering the data, we increased the count rate limits to 0.5 count $\mathrm{s}^{-1}$ (MOS) and 0.65 counts $\mathrm{s}^{-1}$ (pn) for this observation only. For all observations the filtering levels were scrutinized to ensure that periods of significant flaring were entirely removed. When necessary, we made the high-energy count rate thresholds more stringent. 

Point sources were detected using the individual tasks that make up \texttt{edetect\_chain}. This task uses \texttt{eboxdetect} to perform a sliding box detection of sources with a local background, then has \texttt{esplinemap} generate a source-corrected global background, which a second run of \texttt{eboxdetect} uses to find sources again. Sources were selected from these detections by hand after a visual inspection. Table \ref{tab:masked_source} lists the coordinates and radii of a sample of the sources; a full table is provided in the online edition. All identified sources were excluded from the spectral extraction regions.

We then extracted spectra from the observations in three different apertures using standard options (\texttt{"\#XMMEA\_EM"} for MOS data and  \texttt{"\#XMMEA\_EP"} for pn). For all cameras we selected single and double events, filtering with \texttt{"PATTERN \textless 12"} for MOS and \texttt{"PATTERN \textless 4"} for pn. Our first aperture was 300 $h_{70}^{-1}$ kpc, which was chosen to compare our measured fluxes against those of \cite{1998ApJ...502..558V}. Our second aperture was a circle with radius equal to the value of $\mathrm{r}_{2500}$ given in Table 2 of \cite{2011ApJ...726...48H}. $\mathrm{r}_{2500}$ is the radius inside which the estimated mean mass density is 2500 times the critical density at the redshift of the cluster. Weak lensing mass measurements used in this work were derived for $\mathrm{r}_{2500}$ for each cluster. This radius was typically between 40-60$''$. For background regions, we used annuli centered on the cluster with inner radii of 1.2$'$ and outer radii of 1.8$'$. We chose to use this size to obtain as local a background on the detector as possible without any detectable cluster emission present. For typical ranges of beta-profile parameters \citep{1998ApJ...502..558V} we estimate that our choice of background annuli may slightly over-subtract the flux at $<1\%$ level, well below our statistical uncertainties. This estimate is conservative because a single beta-model tends to over-predict the X-ray surface brightness when extrapolated to large radii \citep[e.g.][]{2008MNRAS.387..631E}.

With one exception, when choosing a center for our apertures, we used the Brightest Cluster Galaxy (BCG) coordinates presented in \cite{2011ApJ...726...48H}. This position is the center around which they estimate \rrad\, and \Mrad, and a direct comparison between the mass and X-ray properties of a cluster should be within the same area. The exception, RXJ0826.1+2625, we will show in Section \ref{sect:flux_comp}, is an example where the \textit{ROSAT} center is in error due to point source contamination. \cite{2011ApJ...726...48H} identified a BCG with a reported ``quality'' of the BCG detection of 0, implying an ambiguous identification. Furthermore, their reported value of $\mathrm{M}_{2500} = 0.8_{-2.1}^{+2.1}$ implies a poor determination of the cluster mass around that location. As the center of the X-ray emission detected in \textit{XMM-Newton} is barely within \rrad\, of the reported BCG position, we instead repositioned our aperture around the center of the X-ray emission. The coordinates around which we located our apertures are provided in Table \ref{tab:cluster_properties}. Because of the centering issues, RXJ0826.1+2625 was not included in fits of weak-lensing mass scaling relations.\label{sect:centering}

Spectra were extracted using the SAS task \texttt{evselect}, while redistribution matrix files (RMF) and ancillary response files (ARF) were generated with SAS tasks \texttt{rmfgen} and \texttt{arfgen}, respectively. The task \texttt{backscale} was used to determine the usable area (correcting for bad pixels and CCD edges) for each spectrum. Photon spectra, RMF, and ARF were all binned from 0.4 to 8.0 keV with bins of size 0.038 keV. 

\section{Analysis}
\label{sect:Analysis}
Our extracted spectra were analyzed using XSPEC version 12.8.0 and PyXspec version 1.0.1. For each cluster, three independent spectra from MOS1, MOS2, and pn were fit simultaneously with the same model. In all three observations of RXJ1117.4+0743, the cluster aperture we chose extended outside of the field of view for the MOS2 camera. As this would bias our results toward the properties of the center of the cluster, we did not use those MOS2 data for any of the three observations. Aside from the spectral binning performed in the spectral generation, no binning was performed. Because of that -- and the low number of counts for our objects -- we used the  modified C-statistic \citep{1979ApJ...228..939C,1979ApJ...230..274W} for determining the best fit and uncertainties for our model parameters. 

Our spectra were modeled with a combination of emission (APEC) and absorption (phabs) components from 0.7-8.0 keV. APEC uses the ATOMDB v2.0.2 \footnote{http://atomdb.org/} code to compare the observed data to models of collisionally ionized diffuse gas emission spectra. It requires the redshift \citep[from][]{2003ApJ...594..154M} and metal abundances to fit a normalization and plasma temperature. We used the \texttt{angr} abundance table, which comes from \cite{1989GeCoA..53..197A}.

For all model fits, we used XSPEC to derive flux values from 0.5-2.0 keV, the same range used by \cite{{2003ApJ...594..154M}}. We also calculated luminosities from 0.1-2.4 keV \citep[the range presented in][]{2011ApJ...726...48H}, 0.5-2.0 keV \citep[to match][]{2003ApJ...594..154M}, and 0.1-50 keV (a ``bolometric'' luminosity).
\section{Results}
\label{sect:Results}

\begin{deluxetable*}{lcccccccc}
\tabletypesize{\scriptsize}
\tablecaption{Spectral Fitting Properties Within $\mathrm{r}_{2500}$}
\tablewidth{0pt}
\tablehead{
\colhead{Name} & \colhead{Mass\tablenotemark{a}} & \colhead{kT} & \colhead{Abundance} & \colhead{Norm.} & \colhead{F\tablenotemark{b}} &
\colhead{L\tablenotemark{c}} & \colhead{L\tablenotemark{b}} & \colhead{L\tablenotemark{d}}\\
\colhead{ } & \colhead{$h_{70}^{-1} 10^{13}$} & \colhead{} & \colhead{} & {$10^{-4}$} & 
\colhead{$10^{-14}$} & \colhead{$h_{70}^{-2} 10^{44}$} &
\colhead{$h_{70}^{-2} 10^{44}$} & \colhead{$h_{70}^{-2} 10^{44}$}\\
\colhead{ } & \colhead{$\mathrm{M}_\odot$} & \colhead{keV} & \colhead{$\mathrm{Z}_\odot$} & { APEC\tablenotemark{e} } & 
\colhead{$\mathrm{erg}\,\mathrm{s}^{-1}\,\mathrm{cm}^{-2}$} & \colhead{$\mathrm{erg}\,\mathrm{s}^{-1}$} &
\colhead{$\mathrm{erg}\,\mathrm{s}^{-1}$} & \colhead{$\mathrm{erg}\,\mathrm{s}^{-1}$}
}
\startdata

RXJ0056.9$-$2740 & $ 5.2_{-3.0}^{+4.2}$ & $3.51_{-0.49}^{+0.91}$ & $< 1.03$\tablenotemark{f} & $1.97_{-0.19}^{+0.16}$ & $4.59_{-0.36}^{+0.26}$ & $0.524_{-0.032}^{+0.044}$ & $0.437_{-0.031}^{+0.036}$ & $1.05_{-0.10  }^{+0.08 }$ \\
RXJ0110.3+1938 & $ 5.0_{-2.6}^{+3.4}$ & $2.95_{-0.62}^{+0.72}$ & $0.56_{-0.32}^{+0.50}$ & $0.98_{-0.16}^{+0.19}$ & $3.64_{-0.08}^{+0.20}$ & $0.119_{-0.006}^{+0.003}$ & $0.102_{-0.006}^{+0.002}$ & $0.219_{-0.015}^{+0.016}$ \\
RXJ0522.2$-$3625 & $ 7.2_{-3.1}^{+4.2}$ & $5.32_{-0.37}^{+0.42}$ & $0.37_{-0.12}^{+0.13}$ & $2.11_{-0.09}^{+0.08}$ & $6.11_{-0.10}^{+0.13}$ & $0.454_{-0.012}^{+0.012}$ & $0.371_{-0.009}^{+0.010}$ & $1.19_{-0.03  }^{+0.04 }$ \\
RXJ0826.1+2625 & $ 0.8_{-2.1}^{+2.1}$ & $1.52_{-0.27}^{+0.20}$ & $0.13_{-0.08}^{+0.12}$ & $0.31_{-0.05}^{+0.06}$ & $0.80_{-0.02}^{+0.05}$ & $0.035_{-0.002}^{+0.002}$ & $0.031_{-0.002}^{+0.001}$ & $0.045_{-0.003}^{+0.002}$ \\
RXJ0847.1+3449 & $24.2_{-7.6}^{+8.9}$ & $4.17_{-0.40}^{+0.59}$ & $0.29_{-0.16}^{+0.18}$ & $2.02_{-0.13}^{+0.13}$ & $5.20_{-0.10}^{+0.11}$ & $0.568_{-0.012}^{+0.012}$ & $0.467_{-0.009}^{+0.009}$ & $1.31_{-0.05  }^{+0.03 }$ \\
RXJ0957.8+6534 & $ 4.3_{-2.6}^{+3.2}$ & $2.88_{-0.17}^{+0.21}$ & $0.23_{-0.08}^{+0.10}$ & $1.65_{-0.09}^{+0.09}$ & $3.95_{-0.07}^{+0.09}$ & $0.393_{-0.011}^{+0.009}$ & $0.330_{-0.009}^{+0.009}$ & $0.745_{-0.019}^{+0.012}$ \\
RXJ1117.4+0743 & $ 5.2_{-2.8}^{+3.4}$ & $4.31_{-0.39}^{+0.69}$ & $0.40_{-0.19}^{+0.19}$ & $1.01_{-0.06}^{+0.07}$ & $2.90_{-0.04}^{+0.06}$ & $0.224_{-0.004}^{+0.004}$ & $0.184_{-0.003}^{+0.004}$ & $0.527_{-0.015}^{+0.012}$ \\
RXJ1354.2$-$0221 & $20.2_{-5.6}^{+6.4}$ & $7.55_{-1.21}^{+1.86}$ & $0.38_{-0.27}^{+0.34}$ & $2.55_{-0.19}^{+0.18}$ & $6.89_{-0.19}^{+0.25}$ & $0.679_{-0.020}^{+0.025}$ & $0.547_{-0.024}^{+0.021}$ & $2.12_{-0.09  }^{+0.09 }$ \\
RXJ1642.6+3935 & $ 2.8_{-1.8}^{+2.8}$ & $3.01_{-0.38}^{+0.41}$ & $0.43_{-0.20}^{+0.26}$ & $0.95_{-0.10}^{+0.10}$ & $3.43_{-0.08}^{+0.16}$ & $0.147_{-0.006}^{+0.005}$ & $0.127_{-0.005}^{+0.004}$ & $0.264_{-0.012}^{+0.011}$ \\
RXJ2059.9$-$4245 & $ 4.4_{-2.4}^{+3.3}$ & $2.58_{-0.10}^{+0.10}$ & $0.53_{-0.08}^{+0.10}$ & $1.93_{-0.09}^{+0.09}$ & $7.25_{-0.09}^{+0.12}$ & $0.250_{-0.005}^{+0.003}$ & $0.216_{-0.004}^{+0.004}$ & $0.424_{-0.008}^{+0.007}$ \\
RXJ2108.8$-$0516 & $ 1.8_{-1.4}^{+2.2}$ & $2.34_{-0.49}^{+0.90}$ & $< 2.67$\tablenotemark{f} & $1.16_{-0.29}^{+0.10}$ & $2.89_{-0.32}^{+0.13}$ & $0.097_{-0.008}^{+0.009}$ & $0.082_{-0.003}^{+0.009}$ & $0.161_{-0.014}^{+0.018}$ \\
RXJ2139.9$-$4305 & $ 5.3_{-2.6}^{+3.7}$ & $3.06_{-0.22}^{+0.23}$ & $0.32_{-0.10}^{+0.12}$ & $1.86_{-0.10}^{+0.10}$ & $6.13_{-0.16}^{+0.11}$ & $0.297_{-0.008}^{+0.006}$ & $0.255_{-0.006}^{+0.005}$ & $0.542_{-0.013}^{+0.010}$ \\
RXJ2146.0+0423 & $21.0_{-5.7}^{+6.7}$ & $5.02_{-0.38}^{+0.41}$ & $0.41_{-0.12}^{+0.14}$ & $2.98_{-0.13}^{+0.13}$ & $7.87_{-0.17}^{+0.18}$ & $0.739_{-0.021}^{+0.017}$ & $0.601_{-0.020}^{+0.018}$ & $1.97_{-0.04  }^{+0.06 }$ \\
RXJ2202.7$-$1902 & $ 0.8_{-0.8}^{+2.0}$ & $3.91_{-0.63}^{+0.79}$ & $0.77_{-0.39}^{+0.59}$ & $0.37_{-0.06}^{+0.06}$ & $1.29_{-0.04}^{+0.07}$ & $0.084_{-0.005}^{+0.005}$ & $0.071_{-0.005}^{+0.004}$ & $0.186_{-0.012}^{+0.008}$ \\
RXJ2328.8+1453 & $ 4.0_{-2.6}^{+3.7}$ & $3.12_{-0.23}^{+0.28}$ & $0.38_{-0.12}^{+0.16}$ & $0.59_{-0.04}^{+0.04}$ & $1.58_{-0.02}^{+0.04}$ & $0.135_{-0.003}^{+0.003}$ & $0.113_{-0.003}^{+0.003}$ & $0.269_{-0.005}^{+0.007}$ \\
\enddata
\tablenotetext{a}{Weak lensing masses from \cite{2011ApJ...726...48H}.}
\tablenotetext{b}{0.5 - 2.0 keV.}
\tablenotetext{c}{0.1 - 2.4 keV.}
\tablenotetext{d}{Bolometric.}
\tablenotetext{e}{ $10^{-14} \left( 4\pi[\mathrm{D}_\mathrm{A}(1+z)]^2\right)^{-1} \int n_e n_H dV.$ $\mathrm{D}_\mathrm{A}$ has units cm. $n_e$ and $n_H$ have units $\mathrm{cm}^{-3}.$}
\tablenotetext{f}{$3\sigma$ upper limit.}
\label{tab:results}
\end{deluxetable*}

The results of our spectral fitting are summarized in Table \ref{tab:results}. Mass estimates based on weak-lensing analyses are those reported in \citet{2011ApJ...726...48H}. Our reported luminosities are the unabsorbed luminosities. For all measurements, the reported errors are at the $1\sigma$ level.
\subsection{Flux}
\begin{figure}
\epsscale{1.4}
\plotone{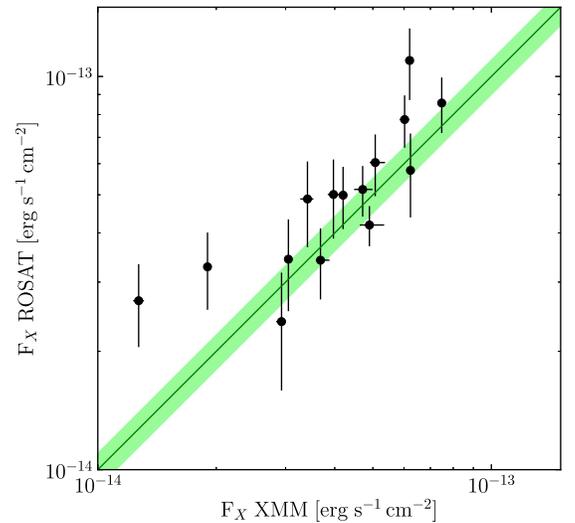}
\caption{Comparison between our measured fluxes using \textit{XMM-Newton} and those reported by \cite{1998ApJ...502..558V} using \textit{ROSAT}. \textit{ROSAT} fluxes were adjusted to correspond to the inner 300 $\mathrm{h}_{70}^{-1}$ kpc of the cluster, as described in the text. The solid line is the identity line, while the shaded band indicates agreement to within 10\%.}
\label{fig:flux}
\end{figure}
\label{sect:flux_comp}

One of our aims was to investigate how improved \textit{XMM-Newton} imaging would affect the measurements of these faint clusters. Along with improved spectral response and calibrations, the improved resolution allowed us to identify and mask out contaminating point sources. To this end, we compare our measured fluxes to those reported in the initial 160SD paper of \citet{1998ApJ...502..558V}, V98 hereafter.

In the original work, V98  were unable to use a wide aperture to integrate flux due to the large statistical uncertainty introduced by the \textit{ROSAT} background. Instead, they estimated the flux from the normalization of a $\beta$-model \citep{1976AA....49..137C},
\begin{equation}
I(r,r_c) = I_0 (1 + r^2/r_c^2)^{-3\beta + 0.5}.
\end{equation}
They estimated core radii by fitting a $\beta$ = 0.67 model to their surface brightness profiles; then, they extrapolated to obtain the flux based on the normalization and shape of the best-fit $\beta$-model. Their final reported flux was actually $(f_{0.6} + f_{0.7})/2$, where $f_{0.6}$ and $f_{0.7}$ are the fluxes obtained assuming $\beta = 0.6$ and $\beta = 0.7$, respectively.

For direct comparison with these results, we integrated counts inside a fixed aperture. In order to avoid biasing these results by our somewhat uncertain estimation of \rrad, we adopted a metric aperture of radius 300 $h_{70}^{-1}$ kpc. For the equivalent flux, we used the $\beta-$model parameters from V98 to infer the estimated \textit{ROSAT} fluxes inside 300 $h_{70}^{-1}$ kpc. The errors on these fluxes were kept at the same percent as the originally reported values.  Details of this procedure are given in Appendix \ref{apx:flux_conversion}. The comparison between our results and V98 is shown in Figure \ref{fig:flux}.

Our measured fluxes agree to within 1$\sigma$ with the modified fluxes of V98 in all but six cases. For RXJ0847.1+3449, including an \textit{XMM-Newton} point source blended with the cluster causes the measured fluxes to agree within their combined 1$\sigma$ errors. To match our flux measurement of RXJ0056.9$-$2740 with that of V98, we only needed to center our aperture on the same position. RXJ2146.0+0423, which we find to be slightly lower in flux than allowed by V98's uncertainty, matches perfectly when we shift to the V98 coordinates and expand the aperture to include a nearby \textit{XMM-Newton} point source. To account for our expanded aperture, we rederived a new, corrected V98 flux to compare in this case. Similarly, repositioning our aperture around RXJ0522.2$-$3625 and using a larger aperture brings the two measurements into agreement. Finally, RXJ0826.1+2625 and RXJ2328.8+1453 were originally  measured at a significant positional offset from V98 ($\approx 37''$ and 45$''$, respectively). In both cases, it appears as if the \textit{ROSAT} images blended in nearby point sources. By recentering our aperture around the V98 coordinates and expanding the region to include the neighboring objects, we find agreement between the two sets of flux measurements. 

We have reproduced the \textit{ROSAT} X-ray flux estimates from V98 and demonstrated that blended point sources and off-center apertures affected the flux estimates of these clusters over and above the uncertainty based on counting statistics and background subtraction alone.

\subsection{X-ray Offsets}
\label{sect:xrayoffsets}
\begin{figure}
\epsscale{1.1}
\plotone{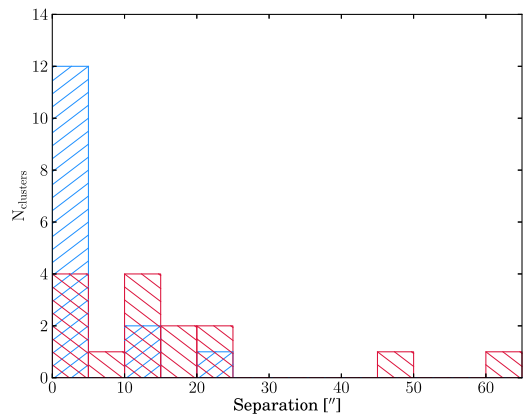}
\caption{Distribution of X-ray centroid offsets to BCG positions measured by \cite{2011ApJ...726...48H} using \textit{XMM-Newton} (this work, hashes rising to the right) and \textit{ROSAT} \citep[][hashes lowering to the right]{1998ApJ...502..558V}. Data are binned to increments of 5 arcseconds.}
\label{fig:pos_hist}
\end{figure}

In Table \ref{tab:cluster_properties} we list coordinates for each cluster twice. The coordinates from \cite{2011ApJ...726...48H} are their best estimate of the position of each cluster's BCG. The new coordinates are of an X-ray centroid performed on data from the MOS1 camera around each cluster's X-ray emission. Centroids were computed in five iterations of centroiding an aperture with radius 16$''$, of images binned to 1.6$''$ per pixel. Twelve of the X-ray-determined positions are within 5$''$ of the BCG position, and the only position more than 12$''$ from the BCG is for RXJ0826.1+2625, where the BCG identification may be questionable. We plot our results, along with the offsets using X-ray positions from \textit{ROSAT}, in Figure \ref{fig:pos_hist}. These \textit{XMM-Newton} observations provide a significant improvement in the ability to properly detect the cluster position over the original \textit{ROSAT} detection positions.

\subsection{Scaling Relations}
\begin{deluxetable*}{lcccc}
\tabletypesize{\scriptsize}
\tablecaption{CCCP Cluster Properties Within $\mathrm{r}_{2500}$}
\tablewidth{0pt}
\tablehead{
\colhead{Name} & \colhead{Mass $\times$ E(z) } & \colhead{L/E(z)} &  \colhead{kT} & \colhead{Redshift}\\
\colhead{ } & \colhead{$h_{70}^{-1} 10^{14}\,\mathrm{M}_\odot$} &  \colhead{$h_{70}^{-2} 10^{45}\,\mathrm{erg}\,\mathrm{s}^{-1}$} & \colhead{keV} & \colhead{ }
}
\startdata
3C295 & $3.30 \pm 0.86$ & $0.78 \pm 0.01$ & $6.43 \pm 0.35$ & 0.464\\
Abell0068 & $2.87 \pm 0.645$ & $0.96 \pm 0.02$ & $7.25 \pm 0.34$ & 0.255\\
Abell0115N & $0.68 \pm 0.46$ & $0.47 \pm 0.01$ & $4.84 \pm 0.10$ & 0.197\\
Abell0115S & $0.84 \pm 0.53$ & $0.32 \pm 0.01$ & $5.60 \pm 0.24$ & 0.197\\
Abell0209 & $2.05 \pm 0.43$ & $0.97 \pm 0.01$ & $7.14 \pm 0.34$ & 0.206
\enddata
\tablecomments{Table \ref{tab:cccp} is published in its entirety in the electronic edition of the Astrophysical Journal. A portion is shown here for guidance regarding its form and content.}
\label{tab:cccp}
\end{deluxetable*}

We fit our measurements of bolometric luminosity, temperature, and mass inside $\mathrm{r}_{2500}$ to the relation
\begin{equation} \label{eqn:Y-Xreltn}
\log \left(\frac{\mathrm{Y}}{\mathrm{Y}_0}\right) = \alpha \log\left( \frac{\mathrm{X}}{\mathrm{X}_0}\right) + \mathrm{C}_X  .
\end{equation}
$\mathrm{X}_0$ and $\mathrm{Y}_0$ are pivot values, which were $10^{44}\,\mathrm{erg}\,\mathrm{s}
^{-1}$, 4 keV, and $10^{14} \, \mathrm{M}_\odot$ for luminosity, temperature, and mass, respectively. Luminosity and mass were corrected for redshift evolution by including the factor E(z); fits were therefore of L/E(z) and ME(z). To extend the dynamic range of our sample and to compare our low mass sample with a higher mass sample at similar redshift, we also included data from the Canadian Cluster Comparison Project \citep[hereafter CCCP]{2012MNRAS.427.1298H,2013ApJ...767..116M}. This sample of 50 galaxy clusters spans redshifts 0.15 \textless\, z \textless\, 0.55, and all clusters were required to have a temperature $\mathrm{k}_B\mathrm{T}_X > 3$ keV. CCCP data was acquired through the online database\footnote{http://sfstar.sfsu.edu/cccp}. In an erratum \citep{2014.JACO.Erratum} these data have been updated since original publication to fix an error in the bolometric luminosity correction factor. We therefore present all of the cluster properties used for fitting in Table \ref{tab:cccp};  a full table is provided in the online edition..

\begin{deluxetable*}{ccccccl}
\tabletypesize{\scriptsize}
\tablecaption{Scaling Relations}
\tablewidth{0pt}
\tablehead{
\colhead{X} & \colhead{Y} & \colhead{Sample} & \colhead{Log Slope} & \colhead{Log Intercept} & \colhead{Bootstrapped} & \colhead{Notes} }
\startdata
L/E(z) & ME(z) & CCCP+160SD & $0.305 \pm 0.042$ & $ 0.134 \pm 0.043$ & NO & WLS, $\sigma_{\log (M|L)} = 0.100$ \\
L/E(z) & ME(z) & CCCP+160SD & $0.435 \pm  0.047 $&  $-0.039 \pm 0.049$ & YES & BCES(Y$|$X) \\
L/E(z) & ME(z) & CCCP & $0.291 \pm 0.075$ &  $0.135 \pm 0.082 $ & YES&  WLS, $\sigma_{\log (M|L)} = 0.137 \pm 0.028 $\\
L/E(z) & ME(z) & CCCP & $0.379 \pm 0.081$ &  $0.005 \pm 0.091$ & YES& BCES(Y$|$X) \\
L/E(z) & ME(z) & 160SD & $1.02 \pm 0.17$ & $0.195 \pm 0.076 $ & YES& BCES(Y$|$X)\\
ME(z) & L/E(z) & CCCP+160SD & $2.33 \pm 0.27$& $0.079 \pm 0.111$ & YES & BCES(X$|$Y)\\
ME(z) & L/E(z) & CCCP & $2.78 \pm 0.73$& $-0.071 \pm 0.311$& YES & BCES(X$|$Y)\\
ME(z) & L/E(z) & 160SD & $1.01 \pm 0.225$& $-0.186 \pm 0.066$& YES & BCES(X$|$Y)\\
L/E(z) & T & CCCP+160SD & $0.229  \pm 0.016$ & $0.005 \pm 0.015$ & YES& WLS, $\sigma_{\log (T|L)} =0.073 \pm 0.009$\\
L/E(z) & T & CCCP+160SD & $0.225 \pm 0.016$ & $   0.012 \pm 0.015$ & YES& BCES(Y$|$X) \\
L/E(z) & T & CCCP & $0.257 \pm 0.029 $ & $-0.026 \pm 0.029$ & YES & WLS, $\sigma_{\log (T|L)} =0.070 \pm 0.009$\\
L/E(z) & T & CCCP & $0.261 \pm 0.029$ & $ -0.028 \pm 0.028$ & YES & BCES(Y$|$X) \\
L/E(z) & T & 160SD & $0.300 \pm 0.055 $ & $  0.052 \pm 0.030$ & NO & WLS, $\sigma_{\log (T|L)} = 0.066$\\
L/E(z) & T & 160SD & $0.293 \pm 0.064 $ & $  0.063 \pm 0.039$ & YES & BCES(Y$|$X)\\
T & L/E(z) & CCCP+160SD & $4.47 \pm 0.33$ & $-0.057 \pm 0.072$ & YES & BCES(X$|$Y) \\
T & L/E(z) & CCCP & $3.88 \pm 0.45$ & $0.098 \pm 0.100$ & YES & BCES(X$|$Y) \\
T & L/E(z) & 160SD & $3.29 \pm 0.57$ & $-0.225 \pm 0.090$ & NO & BCES(X$|$Y) \\
T & ME(z) & CCCP+160SD & $ 1.88 \pm  0.21 $ & $  -0.058  \pm 0.049$ & YES & BCES Bisector \\
T & ME(z)& CCCP+160SD & $ 1.93  \pm 0.24 $ & $  -0.066  \pm 0.053$ & YES & BCES Orthogonal \\
T & ME(z) & CCCP & $ 1.65 \pm  0.24 $ & $  -0.005  \pm 0.061$ & YES & BCES Bisector \\
T & ME(z) & CCCP & $ 1.80 \pm  0.33 $ & $  -0.029 \pm  0.077$ & YES & BCES Orthogonal  \\
T & ME(z) &160SD & $ 1.98 \pm  0.92 $ & $  -0.096  \pm 0.100$ & NO & BCES Bisector\\
T & ME(z) &160SD & $ 1.79  \pm 0.96 $ & $  -0.103  \pm 0.101$ & NO & BCES Orthogonal\\
ME(z) & T & CCCP+160SD & $0.537 \pm 0.059$ & $0.029 \pm 0.024$ & YES & BCES Bisector\\
ME(z) & T & CCCP+160SD & $0.525 \pm 0.065$& $0.032 \pm 0.024$& YES & BCES Orthogonal\\
ME(z) & T & CCCP & $0.622 \pm 0.097$&$-0.008 \pm 0.040$ & YES & BCES Bisector\\
ME(z) & T & CCCP & $0.574 \pm 0.111$&$0.009 \pm 0.044$ & YES & BCES Orthogonal\\
ME(z) & T & 160SD & $0.506 \pm 0.235$ & $0.049 \pm 0.066$& NO & BCES Bisector\\
ME(z) & T & 160SD & $0.559 \pm 0.300$ & $0.058 \pm 0.079$& NO & BCES Orthogonal
\enddata
\label{tab:fit}
\end{deluxetable*}

Individual fits are discussed below, but the results are given in Table \ref{tab:fit}. Fits including data in this work are labeled ``160SD,'' while those including CCCP data are marked as such. Except where noted, uncertainties in fit values were derived through 50,000 bootstrap resamplings. Fits were performed using the WLS and BCES methods described by \citet{1996ApJ...470..706A}. Where luminosity was serving as the X variable, we used the WLS and BCES (Y$|$X) methods, which minimized the residuals in the other parameter. Conversely, when luminosity was the Y variable, we used the BCES (X$|$Y) method. When fitting the mass-temperature relation, we used the BCES Bisector and Orthogonal methods, which considers the residuals in both variables. To account for asymmetric error bars, we estimated a single, logarithmic error for a value $\mathrm{X}^{+\mathrm{u}}_{-\mathrm{d}}$ to be
\begin{equation}
\sigma = 0.4343 \frac{0.5(\mathrm{u} + \mathrm{d})}{\mathrm{X}}.
\end{equation}
For clarity, when describing a relation fit by Equation (\ref{eqn:Y-Xreltn}), we call it the Y-X relation, where X is the independent variable.

\begin{figure}
\epsscale{1.0}
\plotone{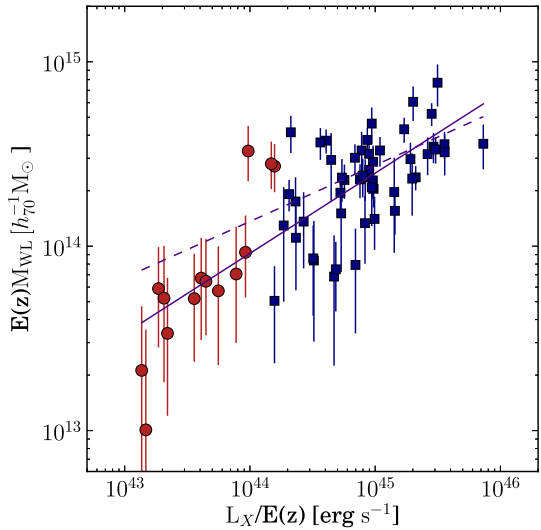}
\caption{Plot of $\mathrm{M}_{WL}$ as a function of bolometric X-ray luminosity within \rrad. Masses and luminosities have been rescaled by E(z) to account for the range of redshift covered by the samples. Data analyzed in this work are shown as circles, while cluster properties from the CCCP are shown as squares. RXJ0826.1+2625 was not included  in this fit. Our best fit to Equation (\ref{eqn:Y-Xreltn}) for the M-L relation is shown by the solid line. Our best fit when including intrinsic scatter is shown by the dashed line.}
\label{fig:mx_lx}
\end{figure}

Our first fit was of the luminosity-mass relation within $\mathrm{r}_{2500}$. When fitting this relation, we did not include RXJ0826.1+2625, as its mass was not well determined (as discussed in Section \ref{sect:centering}). We first fit this relationship without assuming intrinsic scatter; the resulting best-fit slope was $\alpha = 0.435 \pm 0.047$. This result shows no significant difference from the result for the 50 CCCP clusters alone, but it does not agree with the result for a fit only of the low-mass sample presented here. We caution that this discrepancy is not necessarily indicative of a break in the scaling relation, for reasons we will discuss in Section \ref{sect:Discussion}. 

When allowing for intrinsic scatter, the best-fit value of $\alpha$ is $0.305 \pm 0.042$, with an intrinsic scatter of $\sigma_{\log (M|L)} = 0.100$. Figure \ref{fig:mx_lx} shows both fits along with the cluster properties for both samples. For a direct comparison of the reduced scatter, we fit the M-L relation using luminosities from the original work by \cite{2011ApJ...726...48H}. With these, the intrinsic scatter was $\sigma_{\log (M|L)} = 0.262$.  

\begin{figure}
\epsscale{1.0}
\plotone{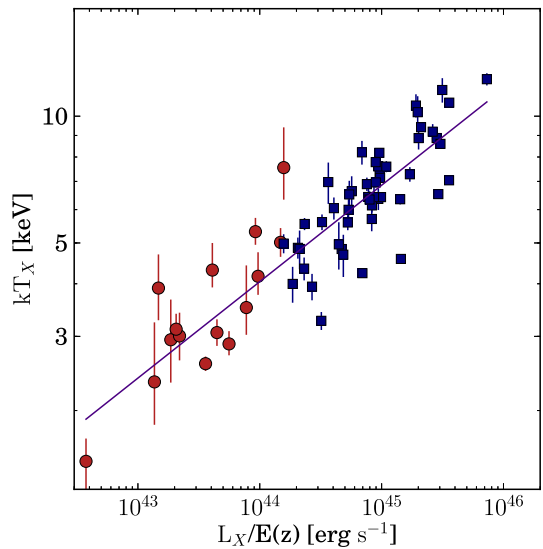}
\caption{Plot of X-ray temperature as a function of bolometric luminosity within \rrad. Luminosities have been rescaled by E(z) to account for the range of redshift covered by the samples. Data analyzed in this work are shown as circles, while cluster properties from the CCCP are shown as squares. Our best fit to Equation (\ref{eqn:Y-Xreltn}) for the T-L relation is shown by the solid line.}
\label{fig:tx_lx}
\end{figure}

Next we fit the temperature-luminosity relation within $\mathrm{r}_{2500}$, this time using all fifteen clusters studied here. We found that the best fit for the entire sample was $\alpha = 0.229  \pm 0.016$ with an intrinsic scatter of $\sigma_{\log (T|L)} = 0.073 \pm 0.009$, consistent with the fits for the two individual samples. This fit is shown along with the data in Figure \ref{fig:tx_lx}. When we did not allow for intrinsic scatter, we found the best-fit slope was relatively unchanged, becoming $\alpha = 0.225  \pm 0.016$.

\begin{figure}
\epsscale{1.0}
\plotone{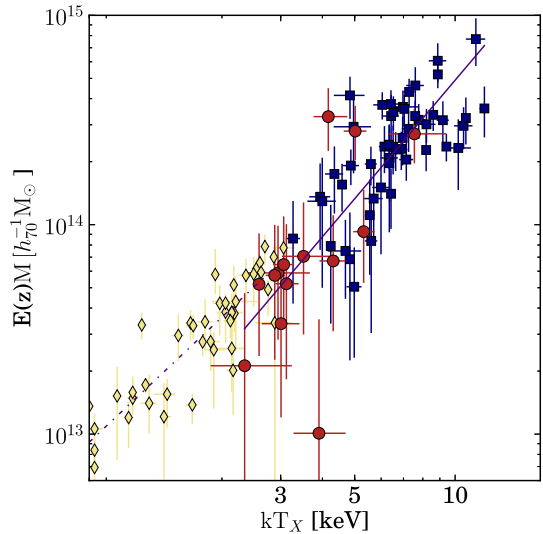}
\caption{Plot of $\mathrm{M}$ as a function of X-ray temperature within \rrad. Masses have been rescaled by E(z) to account for the range of redshift covered by the samples. Data analyzed in this work are shown as circles, while cluster properties from the CCCP are shown as squares; both are derived from weak lensing. RXJ0826.1+2625 was not included  in this fit. We also include a sample of nearby galaxy groups from \citet{2009ApJ...693.1142S} as diamonds, where masses are derived from hydrostatic equilibrium. Our best fit to Equation (\ref{eqn:Y-Xreltn}) for the M-T relation from the clusters analyzed in this work and from the CCCP is shown by the solid line. Our best fit to the M-T relation using the properties within $\mathrm{r}_{2500}$ of the groups from Sun et al. is shown as a dashed line.}
\label{fig:mx_tx}
\end{figure}

We also investigated the scaling between mass and temperature within $\mathrm{r}_{2500}$. Again, RXJ0826.1+2625 was excluded from this fit. For the combined sample, the best-fit with the BCES Bisector was $\alpha = 1.88 \pm 0.21$, which was consistent with fits for the sub-samples alone. This fit is shown in Figure \ref{fig:mx_tx}. In addition, we include data taken from \citet{2009ApJ...693.1142S}. Masses from that study are not based on weak lensing measurements, but were instead derived from an assumption of hydrostatic equilibrium. These data were not included in our fits, however. 

In order to more easily compare our work to other studies, we also fit the inverse of these three relations. Using BCES(X$|$Y), the L-M relation fit for the CCCP+160SD sample is $\alpha = 2.33 \pm 0.27$. In contrast, for BCES(Y$|X$), the inverse of the M-L relation is $\alpha^{-1} = 2.30$. Our BCES(Y$|$X) fit of L-T is $\alpha = 4.47\pm0.33$, while the corresponding fit from the T-L relation is $\alpha^{-1} = 4.36$. When fitting T-M, the best fit from BCES Bisector was $\alpha = 0.537 \pm 0.059$, which agrees with the BCES Bisector of M-T, $\alpha^{-1} = 0.532$.

\section{Discussion}
\label{sect:Discussion}
\subsection{Comparison with Previous X-ray Observations}

We compared our results to previously published individual \textit{XMM-Newton} results for four clusters (RXJ\allowbreak{}0110.3\allowbreak{}+1938, RXJ0847.1+3449, RXJ1117.4+0743, and RXJ1354.2$-$0221). 
To investigate the differences, we replicated the analysis of previous observations, including their aperture sizes and cosmology. We were able to reasonably reproduce
previous results. The discrepancies arising from systematics such as differences in background choices or particle background screening criteria are smaller than the statistical uncertainty. We find that  
any apparent differences between our results for these clusters with previous results arise because of differences in aperture sizes, and rarely, choice of aperture centers. The details of this comparison are reported in Appendix \ref{apx:xmm_comp}.

Our most obvious source of possible discrepancy with previous works is our choice of apertures, which have a radius $\mathrm{r}_{2500}$ motivated by weak-lensing estimates from \cite{2011ApJ...726...48H} that were unavailable to most of the other studies. Another potential source of X-ray temperature discrepancy is the choice of binning spectral data. 
Some previous works binned spectral data to as few as 12 counts per spectral bin. We leave our spectra unbinned and fit with the C-statistic. As this work is focusing on faint clusters, we are limited by low photon counts. If data are binned such that only a few counts are in each bin, each bin will have non-Gaussian behavior. Since the $\chi^2$ statistic is defined for Gaussian-distributed data, it will not be a valid fitting statistic in this case. Alternatively, data can be binned, but doing so potentially degrades spectral resolution. Along with producing better fits for low counts \citep{1989ApJ...342.1207N,2006AA...451..457T}, use of the C-statistic can also avoid biases in the high-count regime \citep{2009ApJ...693..822H}. Use of different thermal models for fitting spectra did not cause major deviations in our results. As we were able to reproduce the earlier results while still using an APEC model, this should not therefore bias our results significantly \citep[see also][]{2005AA...430..385B, 2007AA...462..953M}.

We demonstrated in detail (see Appendix \ref{apx:xmm_comp}) that we can recover results of previous works, which verifies their results and ours. However we caution that the choice of aperture and center affect the estimate of L, T, and M for any cluster, and that results from different analyses cannot be blindly combined.

\subsection{Comparison with Other Scaling Relations}

We have measured scaling relations between weak lensing mass, X-ray luminosity, and temperature for a sample of clusters with mass and luminosity around the cluster/group boundary and at redshifts $0.3 < \mathrm{z} < 0.6$. As we used weak lensing masses and bolometric luminosities and because we only investigated X-ray properties within $\mathrm{r}_{2500}$, no exact comparisons are available for our results. However, we can compare our results to other similar studies, both those focused on local groups and those that include clusters at redshifts similar to what was studied here but more massive than our sample.

Our best fit of the M-L relation within $\mathrm{r}_{2500}$ was, when neglecting intrinsic scatter, $\alpha = 0.435 \pm 0.047$. \cite{2011ApJ...726...48H} fit this relation using almost the same clusters studied here, with \textit{ROSAT} luminosities taken from the \textit{ROSAT} measurements, and a higher mass sample, reporting $\alpha = 0.68 \pm 0.07$. Other works \citep{2007ApJ...668..772M, 2008MNRAS.387L..28R, 2011AA...535A.105E, 2011AA...535A...4R} find values in the range $0.5 \lesssim \alpha \lesssim 0.75$, consistent with but somewhat steeper than ours. 

For the L-T relation within $\mathrm{r}_{2500}$, we found a best fit of $\alpha = 4.47 \pm 0.33$, although the 160SD groups and the CCCP clusters each had shallower slopes when fit independently. Previous results \citep{2007ApJ...668..772M, 2009AA...498..361P, 2010ApJ...724..608B, 2011AA...535A.105E, 2011AA...535A...4R, 2014AA...564A..17N} have reported slopes from $2.5 \lesssim \alpha \lesssim 4.5$. Our results, particularly for the two sub-samples fit individually, are consistent with this range, albeit on the high end.

In this work, we reported the best fit slope of the M-T relation within $\mathrm{r}_{2500}$ was $\alpha = 1.88 \pm 0.21$. Previous works \citep{2009ApJ...693.1142S,2011AA...535A.105E,2011AA...535A...4R, 2013ApJ...778...74K} have reported slopes in the range $1.45 \lesssim \alpha \lesssim 1.85$. As was the case with the T-L relation, our slope for the combined sample is slightly higher than this range, but the group and cluster samples, when fit independently, are both in agreement with these studies.

While comparing our scaling relationships to others is worthwhile, we caution that there are a handful of issues that make direct comparison problematic. As mentioned earlier in this discussion, other works used different radii within which to measure cluster properties. Our choice of $\mathrm{r}_{2500}$ was motivated by the requirements imposed from our weak lensing masses, but it means we are analyzing X-ray properties in different apertures from other studies.

Another issue that arises when comparing to other studies is the definition of luminosity. In this work, we used bolometric luminosities. However, in the three works that looked at groups that we discuss in this section \citep{2008MNRAS.387L..28R,2011ApJ...726...48H,2011AA...535A.105E}, all fit scaling relations with a luminosity only within the energy band of 0.1 - 2.4 keV. The importance of energy bands was shown by \citet{1998ApJ...504...27M}, who found that when switching from luminosities within 0.1-2.4 keV to bolometric luminosites the measured slope of the L-T relation steepened from $\alpha = 2.10 \pm 0.24$ to $\alpha = 2.64 \pm 0.27$. Such a large change in the fit means that we should be careful comparing scaling relations for luminosities derived from different energy bands. As a test of this effect, we fit the M-L and T-L relations using luminosities measured in the 0.1 - 2.4 keV energy band for the 160SD clusters. The power law indices increased when using the energy limited luminosities from $1.02 \pm 0.17$ to $\alpha = 1.19 \pm 0.22$ and from $0.293 \pm 0.064 $ to $\alpha = 0.334 \pm 0.101$ for M-L and T-L, respectively. The 160SD sample here has too small a dynamical range to be seriously considered for a scaling relation, but the effect of choosing to fit bolometric luminosities over band-limited luminosities is clear.

Also, while our sample is a subset of a randomly selected survey, it is originally based on X-ray selected clusters. \cite{2013MNRAS.431.2542H} suggest that X-ray selection preferentially picks centrally concentrated systems; these systems populate the high $\mathrm{L}_X$ side of the T-L relation. Similarly, since our data were drawn from the faint end of a flux-limited sample, we would expect preferentially over-luminous clusters for their mass to be selected.

\subsection{Comparison to Low-Redshift Groups}
One of the issues with comparing the difference between the groups examined in this paper and those at low redshift is the ubiquity of masses derived from hydrostatic equilibrium. Hydrostatic masses may somewhat underestimate the mass of galaxy clusters when compared to weak lensing measurements \citep{2007AA...474L..37A, 2008MNRAS.384.1567M,2013ApJ...767..116M}. However, in order to allow a comparison with work on low redshift groups and poor clusters using hydrostatic masses, we make the assumption that both mass estimates are identical. 

Looking at Figure \ref{fig:mx_tx}, we can see that our moderate redshift clusters ($\overline{z} = 0.444$) are almost all hotter and/or less massive than what would be predicted by the lower redshift scaling relations for groups presented in \citet{2009ApJ...693.1142S} ($\overline{z} = 0.042$), although five low-temperature clusters agree very well. If we test the hypothesis whether our data are fit by the Sun2009 relationship between mass and temperature, we find a $\chi^2$ value of 19.09 for 14 clusters ($p=0.089$). Therefore, to within $2-\sigma$ we see no difference between our sample and the low-redshift sample. If we limit this analysis to only those clusters with temperature kT \textless 4 keV, our value of $\chi^2$ is 3.15 for 9 clusters ($p=0.87$). If we do not scale the mass by E(z), $\chi^2$ becomes 29.53 for 14 clusters ($p=0.0033$). This significance is just below 3$\sigma$, constituting very weak evidence for the expected self-similar evolution in the temperature-mass relation for groups.

\subsection{Comparison between Groups and Clusters}
A direct comparison between the CCCP sample and our sample, which affords a comparison between low-mass and high-mass clusters at a similar redshift range, is difficult due to the limited number of clusters in both sets. So to provide some quantification of whether the two populations differ, we utilize Fisher's exact test, which looks at how two properties are distributed in two populations. In this case, we look at how our sample and the CCCP sample compare to the scaling relations. We choose to use Fisher's exact test due to how few objects we have; in this domain, Fisher's exact test is the best, if not the only, test to use \citep{2012psa..book.....W}.

For both samples, we count how many clusters lie above the lines of best fit for each scaling relation and how many lie below. Our null hypothesis is that the samples are similar and so the number of clusters above the relation should equal the number below, statistically. We compute p-values for the T-L, M-L, and M-T relations of $p=0.13$, $p=0.19$, and $p=0.19$, respectively. We therefore cannot reject the hypothesis that groups and clusters at intermediate redshift behave identically with respect to the scaling relations derived in this work, so our measurements are consistent with the hypothesis that z$\sim$0.3-0.5 X-ray selected clusters and groups/poor clusters follow similar X-ray scaling laws.

\section{Conclusions}
We have presented new and revised X-ray properties for a sample of 15 galaxy clusters originally drawn from a random sample of the 160 Square Degree Survey. Covering a range of redshifts from $0.3 < \mathrm{z} < 0.6$ and limited in mass to $\mathrm{M}_{2500} \lesssim 2 \times 10^{15} \, \mathrm{M}_\odot$, our new X-ray data together with previously published HST weak lensing measurements probe a largely-unexplored parameter space in cluster mass and redshift. By using a rigorous analysis to match cluster properties measured within the same radius as existing weak-lensing masses, we investigate scaling relations between mass, luminosity, and temperature. Our primary conclusions are summarized below.

1. We measure fainter fluxes than reported from earlier \textit{ROSAT} measurements \citep{1998ApJ...502..558V} for five of the fifteen clusters studied here (RXJ0522.2$-$3625, RXJ0826.1+2625, RXJ0847.1+3449, RXJ2146.0+0423, RXJ2328.8+1453). Due to a combination of fainter sources blending into the extended cluster light profile and multiple sources blending into one, we also found that reported X-ray positions for these clusters were not accurate. Due to the original positional inaccuracy, RXJ0056.9$-$2740 was originally reported to be fainter than we measured. Use of {detections near the flux threshold of objects subject to blending because of the angular resolution, such as \textit{ROSAT} cluster surveys,} can therefore lead to errors in both position and flux that can be larger than the quoted statistical flux uncertainty.

2. Inside $\mathrm{r}_{2500}$, for the mass and redshift range studied here, the fourteen clusters with reasonable mass measurements and 50 clusters from the CCCP are best fit by the relation
\begin{equation}
\frac{\mathrm{ME(z)}}{10^{14}\,\mathrm{M}_\odot} = 10^{-0.04 \pm 0.05} \times \left( \frac{\mathrm{LE(z)}^{-1}}{10^{44}\,\mathrm{erg}\,\mathrm{s}^{-1}} \right)^{0.44 \pm 0.05}.
\end{equation} 
When we allow for intrinsic scatter, the exponent of the best fit becomes $0.31 \pm 0.04$. Our results indicate neither a break in the scaling relation among groups nor increased scatter at low mass.

3. Using uncontaminated luminosity measurements and uniformly-defined $\mathrm{r}_{2500}$ values from weak lensing, the intrinsic scatter in the L-M relation reduced from $\sigma_{\log (M|L)} = 0.26$ to $\sigma_{\log (M|L)} = 0.10$. 

4. Similarly, when determining the scaling relation between luminosity and temperature within $\mathrm{r}_{2500}$, we find that 
\begin{equation}
\frac{\mathrm{k}_\mathrm{B}\mathrm{T}}{4\,\mathrm{keV}} = 10^{0.005 \pm 0.015} \times \left( \frac{\mathrm{LE(z)}^{-1}}{10^{44}\,\mathrm{erg}\,\mathrm{s}^{-1}} \right)^{0.23 \pm 0.02}.
\end{equation}
We find a small intrinsic scatter of $\sigma_{\log (\mathrm{T}|\mathrm{L})} = 0.07 \pm 0.01$. When the high- and low-mass samples are fit separately, the 50 clusters from the CCCP sample and the 15 clusters from the 160SD sample scaling relations each have steeper slopes, $0.26 \pm 0.03$ and $0.30 \pm 0.06$, respectively. Again, we find no evidence of a break in this relation among groups.

5. For scaling between weak-lensing masses and X-ray temperatures within $\mathrm{r}_{2500}$, the combined sample of 14 clusters from this work with reasonable mass measurements and 50 from the CCCP are best fit by the relation
\begin{equation}
\frac{\mathrm{ME(z)}}{10^{14}\,\mathrm{M}_\odot} = 10^{-0.06 \pm 0.05} \times \left( \frac{\mathrm{k}_\mathrm{B}\mathrm{T}}{4\,\mathrm{keV}} \right)^{1.9 \pm 0.2}.
\end{equation}
When fitting high and low mass subsamples independently, the slope becomes $1.7 \pm 0.2$ and $1.8 \pm 1.0$, respectively. These results agree with other results for both nearby groups and intermediate-redshift clusters, along with the self-similar prediction that $\mathrm{M} \propto \mathrm{T}^{3/2}$. 

6. To the statistical limits of our data, the intermediate redshift groups are within 2$\sigma$ of the M-T relation extrapolated from a low-redshift group sample from \citet{2009ApJ...693.1142S}. Without self-similar evolution, there is a deviation just below the level of 3$\sigma$, indicating a weak statistical preference for the expected self-similar evolution.\\

The authors wish to thank Seth Bruch for his work planning this project. M.D. and T.C. acknowledge partial support from a NASA ADAP award NNX11AJ60G. A.M. acknowledges support from NASA grant NNX12AE45G. This work is based on observations obtained with \textit{XMM-Newton}, an ESA science mission with instruments and contributions directly funded by ESA Member States and NASA.

{\it Facility:} \facility{XMM}

\appendix
\section{Conversion of ROSAT fluxes to \texorpdfstring{$\mathrm{f}_{300\, \mathrm{kpc}}$}{fluxes inside 300 kpc}}
\label{apx:flux_conversion}
In this appendix, we discuss how we converted the fluxes reported by \cite{1998ApJ...502..558V} into aperture fluxes. As the original fluxes were found by integrating a $\beta$-model to infinity, we derived a means of obtaining the normalization from a given flux. We then integrated the $\beta$-model to a desired angular aperture using this normalization.

The flux of a $\beta$-model is found by integrating the intensity
\begin{equation}
f = \int I_0  \left( 1 + \left( \frac{\theta}{\theta_c} \right)^2 \right)^{-3 \beta + 0.5} 2 \pi \theta d\theta.
\end{equation}
Substituting $x = -3 \beta + 0.5$, this is an analytic integral with solution
\begin{equation}
f = 2 \pi I_0 \frac{(\theta_c^2 + \theta^2)(\theta^2/\theta_c^2 + 1)^x}{2(x+1)} + c
\end{equation}
When evaluating this as $\theta \rightarrow \infty$ for any $x < 1$, the upper part of the fraction will go to 0. When evaluating at $\theta = 0$, this becomes 
\begin{equation}
\label{eqn:f_at_0}
f(\theta=0) = 2 \pi I_0 \frac{\theta_c^2}{2(x+1)}.
\end{equation}
As \cite{1998ApJ...502..558V} reported their fluxes as the average of the fluxes found with $\beta = 0.6$ and $\beta = 0.7$, we can determine their normalization, $I_0$, by inserting the appropriate values of x and rearranging Equation (\ref{eqn:f_at_0}). We use $x_{0.6}$ and $x_{0.7}$ to denote the values of x found with $\beta = 0.6$ and $\beta = 0.7$, respectively, and include a factor of 1/2 to account for averaging, so that we have
\begin{equation}
I_0 = \frac{-2 f_{ROSAT}}{\pi \theta_c^2} \left( \frac{1}{x_{0.6}+1} + \frac{1}{x_{0.7}+1} \right)^{-1}.
\end{equation}
From this, the total flux that would be measured inside a aperture of radius $\theta$ can be computed for a given value of $\beta$ using
\begin{equation}
f_x(\theta) = \frac{-2 f_{ROSAT}}{\theta_c^2(x + 1)}\left( \frac{1}{x_{0.6}+1} + \frac{1}{x_{0.7}+1} \right)^{-1} \left[ (\theta_c^2 + \theta^2)\left( \frac{\theta^2}{\theta_c^2} + 1 \right)^x - \theta_c^2 \right].
\end{equation}
To compare the \textit{ROSAT} fluxes to our own, we solve this for the angle equivalent to 300 kpc for $\beta = 0.6$ and $\beta = 0.7$, averaging the two results.

\section{Replication of Previous XMM Analyses}
\label{apx:xmm_comp}
\subsection{RXJ0110.3+1938}
\cite{2010ApJ...724..608B} first analyzed this cluster with the same observation used in this paper. While their analysis followed a similar path to our own, their reported results are not the same as ours. Our reported bolometric luminosity is similar to theirs ($2.19_{-0.14}^{+0.12}$ and $2.08^{+0.22}_{-0.22}$ $\times$ $10^{43}$ erg $\mathrm{s}^{-1}$, respectively), but their reported temperature is noticeably lower than our own ($1.46^{+0.26}_{-0.19}$ keV compared to $2.95_{-0.62}^{+0.72}$ keV). The difference in the result may arise from their less stringent cut for selecting good time intervals, their grouping of their data into energy bins, their use of a smaller aperture, and their lack of pn observations, which supply around 50\% of the counts but were often problematic to calibrate 5 years ago. If we also make these choices, we measure a new temperature of $1.27_{-0.11}^{+0.06}$ keV, which agrees with the earlier result.

However, when we reduce our aperture size and bin the spectral data, we find an even lower luminosity; our new bolometric luminosity is $0.79_{-0.05}^{+0.04} \times$ $10^{43}$ erg $\mathrm{s}^{-1}$. After private communication with S. Bruch, we discovered that the same spectral fitting results were obtained but not published for an aperture of 0.5 Mpc. Using the 4.647 kpc $\mathrm{arcsec}^{-1}$ scale provided in the refereed paper, we extract spectra from a 107.60$''$ aperture. When letting the abundance vary, we find $\mathrm{T}_X = 1.50_{-0.32}^{+0.45}$ keV and $\mathrm{L}_\mathrm{bolo} = 1.83_{-0.19}^{+0.10} \times$ $10^{43}$ erg $\mathrm{s}^{-1}$. In addition, we find 252 and 219 net counts for MOS1 and MOS2, respectively. These results are in agreement with the earlier result, which found 231 and 205 counts for the two cameras. We therefore conclude that their reported X-ray aperture radius of 32$''$ is incorrectly reported, and the actual aperture used was 0.5 Mpc. Using this aperture, we obtain similar results.

\subsection{RXJ0847.1+3449}
\cite{2004AA...420..853L} originally looked at RXJ0847.1+3449 using \textit{XMM-Newton} observation 0107860501. They reported higher values for flux and bolometric luminosity, but a cooler temperature. One source of this difference may be the larger spectral extraction area they used -- it was 120$''$, while ours was $\approx 70''$. Therefore we attempted to reproduce their results by using the same aperture and masks, as that work included images of where point sources were excluded.

Bolometric luminosities reported by \cite{2004AA...420..853L} are not for the 120$''$ apertures. Rather, they are for apertures scaled to the entire virial radius, as found by using the fitted temperatures and the T -- $\mathrm{r}_v$ relation of \cite{1996ApJ...469..494E}. In addition, they increased the estimated photon count rate to account for lack of spatial coverage due to chip gaps or masked point sources. We find a comparable luminosity by fitting a MEKAL model to the parameters specified in Table 5 of \cite{2004AA...420..853L}. Unlike the reported luminosity, these parameters are for the best fit of the spectrum within 120$''$ and are the best measure of what a similar aperture luminosity would be from that work. In order to allow for changes in MEKAL over the past ten years, we let the abundance vary but match the flux reported in the original work. 

When fitting to data from the larger aperture, our temperature estimate changes from $4.16_{-0.39}^{+0.58}$ keV to  $3.72_{-0.41}^{+0.51}$ keV, which agrees with the reported value of $3.62^{+0.58}_{-0.51}$ keV. Similarly, our flux estimate changes from $5.20_{-0.14}^{+0.12} \times$ $10^{-14}$ erg $\mathrm{s}^{-1} \,\mathrm{cm}^{-2}$ to $6.77_{-0.12}^{+0.14} \times$ $10^{-14}$ erg $\mathrm{s}^{-1} \,\mathrm{cm}^{-2}$, in agreement with the predicted $7.04 \pm 0.3 \times$ $10^{-14}$ erg $\mathrm{s}^{-1} \,\mathrm{cm}^{-2}$. For bolometric luminosity, our value within $\mathrm{r}_{2500}$ is $ 1.31_{-0.03 }^{+0.04 } \times 10^{44}\,h^{-2}_{70} \,\mathrm{erg} \, \mathrm{s}^{-1}$, while inside a $120''$ aperture it is $1.70_{-0.05}^{+0.06}\times 10^{44}\,h^{-2}_{70}$ erg $\mathrm{s}^{-1}$. The expected luminosity inside that aperture is $1.75\times 10^{44}\,h^{-2}_{70}$ erg $\mathrm{s}^{-1}$. 

\subsection{RXJ1354.2$-$0221}

RXJ1354.2$-$0221 was also originally investigated by \cite{2004AA...420..853L}, and, as before, they find a higher flux, higher luminosity, and a lower temperature than we do. As with RXJ0847.1+3449, their technique deviated in aperture size, binning, and definition of luminosity. Additionally, we filtered this data for intervals of flaring differently than they did, which we adjust for in our reanalysis. 

We again find a drop in temperature, which changes from $7.60_{-1.22}^{+1.92}$ keV to $3.88_{-0.59}^{+0.93}$ keV when expanding the aperture, in comparison to the originally reported value of $3.66^{+0.6}_{-0.5}$ keV. Likewise, the flux increases from $6.90_{-0.19}^{+0.15} \times$ $10^{-14}$ erg $\mathrm{s}^{-1} \,\mathrm{cm}^{-2}$ to $10.17_{-0.22}^{+0.18} \times$ $10^{-14}$ erg $\mathrm{s}^{-1} \,\mathrm{cm}^{-2}$, which matches the earlier result of $9.8 \pm 0.5 \times$ $10^{-14}$ erg $\mathrm{s}^{-1} \,\mathrm{cm}^{-2}$. Finally, our luminosity rises from $2.11_{-0.12}^{+0.10} \times 10^{44} \,h^{-2}_{70}$ erg $\mathrm{s}^{-1}$ to $2.47_{-0.06}^{+0.09}\times 10^{44} \,h^{-2}_{70}$ erg $\mathrm{s}^{-1}$, which agrees with the predicted expectation of $2.41\times 10^{44} \,h^{-2}_{70}$ erg $\mathrm{s}^{-1}$. As before, we are able to reproduce the earlier results.

\subsection{RXJ1117.4+0743}
\cite{2007ApJ...664..777C} used the same observations analyzed here to look at RXJ1117.4+0743. Their reported temperature ($3.3^{+0.7}_{-0.6}$ keV) is slightly lower than our own ($4.30_{-0.38}^{+0.70}$ keV), but they find larger luminosities from 0.5-2.0 keV (4.19 $\pm$ 0.35 to our $1.84_{-0.03}^{+0.03}$, in units of $10^{43}$ erg $\mathrm{s}^{-1}$) and in a bolometric band (11.8 $\pm$ 0.9 to our $ 5.27_{-0.16}^{+0.08}$ in units of $10^{43}$ erg $\mathrm{s}^{-1}$). There are a few differences in our analysis that can bring those results into closer alignment. Along with using a larger aperture -- 66$''$ to our choice of 47$''$ -- the previous work binned its data to a minimum of 12 counts per energy bin. Making those adjustments is not enough to match the previous work, however, without also using a different background. In the initial paper, the background was described only as ``a larger extraction region near the detector border without any visible sources." To that end, we used a background centered around $\alpha_{2000}=11^\mathrm{h}17^\mathrm{m}40^\mathrm{s}$, $\delta_{2000}=+07^\circ55^\mathrm{m}10^\mathrm{s}$ that was 72$''$ in size. With this background, we recover similar results to the original reporting: $\mathrm{T}_X = 3.13_{-0.29}^{+0.30}$ keV, $\mathrm{F}_{[0.5 - 2.0 \,\mathrm{keV}]} = 5.29_{-0.13}^{+0.12}\times 10^{-14}$ erg $\mathrm{s}^{-1} \, \mathrm{cm}^{-2}$, $ \mathrm{L}_{[0.5 - 2.0 \,\mathrm{keV}]}= 3.90_{-0.14}^{+0.18}\times 10^{43}$ erg $\mathrm{s}^{-1}$, and $\mathrm{L}_{\mathrm{bolo}} = 8.90_{-0.37}^{+0.47} \times 10^{43}$ erg $\mathrm{s}^{-1}$. Even without knowing their exact background region, we reproduce the results of \cite{2007ApJ...664..777C}.

\end{document}